\documentclass[final]{IEEEtran}
\usepackage{amsthm,amssymb,graphicx,multirow,amsmath,color,amsfonts}
\usepackage[update,prepend]{epstopdf}
\usepackage[noadjust]{cite}
\usepackage[latin1, utf8]{inputenc}
\usepackage{tikz}
\usepackage{bbm} 
\usepackage{pdfpages}
\usepackage{tabulary}
\usepackage{multirow}
\usepackage{comment}
\usepackage{subfigure}
\usepackage{float}
\usepackage{cuted}

\usepackage[titlenumbered,ruled]{algorithm2e}
\usepackage[noend]{algpseudocode}
\DeclareUnicodeCharacter{0177}{\^y}

\def\nb0{{\mathbf{0}}}
\def\nb1{{\mathbf{1}}}

\newtheorem{lemma}{Lemma}

\newtheorem{definition}{Definition}

\newtheorem{theorem}{Theorem}

\newtheorem{remark}{Remark}

\allowdisplaybreaks 
\begin{document}
\title{On the Peak AoI of UAV-assisted IoT Networks: A Stochastic Geometry Approach}
\author{
	Yujie Qin, Mustafa A. Kishk, {\em Member, IEEE}, and Mohamed-Slim Alouini, {\em Fellow, IEEE}
	\thanks{Yujie Qin and Mohamed-Slim Alouini are with Computer, Electrical and Mathematical Sciences and Engineering (CEMSE) Division, King Abdullah University of Science and Technology (KAUST), Thuwal, 23955-6900, Saudi Arabia. Mustafa Kishk is with the Department of Electronic Engineering, Maynooth university, Maynooth, W23 F2H6, Ireland. (e-mail: yujie.qin@kaust.edu.sa; mustafa.kishk@mu.ie; slim.alouini@kaust.edu.sa).} 
	
}
\maketitle
\begin{abstract}
	In this paper, we analyze the peak age of information (PAoI) in UAV-assisted internet of thing (IoT) networks, in which the locations of IoT devices are modeled by a  Mat\'{e}rn cluster process (MCP) and UAVs are deployed at the cluster centers to collect the status updates from the devices. Specifically, we consider that IoT devices can either monitor the same physical process or different physical processes and UAVs split their resources, time or bandwidth, to serve the devices to avoid inter-cluster interference. Using tools from stochastic geometry, we are able to compute the mean activity probability of IoT devices and the conditional success probability of an individual device. We then use tools from queuing theory to compute the  PAoI under two load models and two scenarios for devices, respectively. Our numerical results show interesting system insights. We first show that for a low data arrival rate, increasing the number of correlated devices can improve the PAoI for both load models. Next, we show that even though the time-splitting technique causes higher interference, it has a limited impact on the mean PAoI, and the mean PAoI benefits more from the time-splitting technique. This is because of the nature of UAV communication, especially at places where devices (users) are spatially-clustered: shorter transmission distances and better communication channels, comparing the links established by the cluster UAV and serving devices (users) to links established by interferers.
\end{abstract}
\begin{IEEEkeywords}
	Stochastic geometry, peak age of information,  Mat\'{e}rn cluster process, resource allocation, correlated and uncorrelated device
\end{IEEEkeywords}
\section{Introduction}
With the development of advanced wireless communication and sensing techniques, a massive number of IoT devices are deployed widely to enable multiple real-time applications such as localization, remote monitoring, image collection, and environment sensing. In such real-time involved scenarios, low latency and reliable communication channels are of vital importance. For instance, internet of thing (IoT) devices can be used in disaster monitoring. During the course of any disasters, accurate and up-to-date information help in making timely decisions which can reduce the damage dramatically \cite{8698814,hristidis2010survey}. To this objective, the information age, which characterizes the freshness of the data, draws great attention \cite{kaul2012real,alberts1997information}.  Compared to traditional metrics like delay, the concept of `age of information' (AoI) offers a more insightful perspective. AoI is defined as the time elapsed since the last successfully received update packet at the monitor was generated at the source, effectively capturing the timelines of updates \cite{emara2020spatiotemporal}. As new updates are received, the AoI grows until the arrival of the next update. Therefore, it computes the time duration between two successfully transmitted updates. Consequently,  collecting data from IoT devices efficiently and reducing AoI at the receiver side become crucial to maintain the functionality and high quality of service of the network.

Featured by high flexibility of adjusting locations, and mobility of tracking serving users, unmanned aerial vehicles (UAVs), also known as drones, are expected to be widely acting as communication relays in future wireless networks \cite{mozaffari2019tutorial,li2018uav,zeng2016wireless}. For instance, UAVs can adjust their altitudes to establish line-of-sight (LoS) links with ground users and modify their locations when the spatial distribution of ground users changes \cite{9153823,9444343}. Besides, UAVs are expected to play a key role in IoT networks. Firstly, UAVs can wirelessly charge and wake up IoT devices, which prolongs the lifetime of these energy-limited devices \cite{8967197,9321340,8894454}. Secondly, UAVs can modify their trajectory to get closer and serve the IoT devices based on demands \cite{8842600,9123495}. It is both time and energy efficient to use UAVs as communication relays between IoT devices and BSs. Moreover, at places where the devices are spatially-clustered, such as industrial areas and remote areas, UAVs are more suitable owing to their altitudes, hence, higher LoS probabilities, and better channel quality \cite{Mbref}, compared to the ground BSs. 

 Motivated by the necessity for timely updates in IoT devices and recognizing the benefits of UAV communications, our study focuses on analyzing the PAoI in UAV-assisted IoT networks. In this context, PAoI is defined as the maximum value of AoI immediately before an update packet is received at the monitor, signifying the worst-case scenario for AoI. By investigating the PAoI, we gain insights into the worst AoI situations, which are crucial for ensuring efficient and timely data collection and transmission in UAV-assisted IoT architectures. Specifically, we use tools from stochastic geometry to model the locations of IoT devices, which are spatially distributed according to a Mat\'{e}rn cluster process (MCP), and assume that UAVs hover above the cluster centers to collect information. Additionally, since we consider a cluster of devices, they can either monitor the same physical process or different physical processes, and UAVs need to split their resources to serve them and avoid inter-cluster interference. Hence, in this work, we compute the mean PAoI of UAV-assisted IoT networks under IoT devices in two scenarios: (i) correlated devices and (ii) uncorrelated devices, and two load models: (i) time splitting and (ii) bandwidth splitting. 

\subsection{Related Work}
Literature related to this work can be categorized into: (i) AoI analysis, (ii) stochastic geometry-based approach to the AoI analysis, and (iii)  Analysis of UAV-assisted IoT networks. A brief discussion on related works is provided in the following lines.

 {\em Queueing theory-based AoI analysis.}  An introduction and survey about AoI was provided in \cite{yates2021age}, in which the authors introduced some basic methods for analyzing AoI and mainly focused on the recent contributions in AoI areas. Authors in \cite{8930830} provided an introduction about AoI and its variations, and compared this performance metric to other well-known performance metrics. In \cite{kaul2012real,xiong2004deriving}, authors introduced the concept of the freshness of the data and started to characterize the AoI. Followed by them, AoI analysis has been widely investigated in the literature. For instance, authors in \cite{kaul2012status} studied the AoI based on the queuing discipline of last-come-first-serve. Authors in \cite{soysal2019age} studied the AoI of a G/G/1/1 queuing system and authors in \cite{inoue2019general} derived a general formula for the stationary distribution of the AoI, which is given in terms of system delay and PAoI. Besides, PAoI in an M/M/1 queue and package delivery error was computed in \cite{chen2016age} and mean AoI under M/M/1 queuing system with fixed or random deadline was studied in \cite{8323423}. Some other related works can be found in \cite{champati2019distribution,kosta2021age,zou2019waiting,yates2017status,zhou2020age}.

{\em Stochastic geometry-based approach to the AoI analysis.} The first work of combining AoI and stochastic geometry is provided in \cite{yang2020optimizing}, in which the authors considered a Poisson bipolar network and optimized the AoI by scheduling. Besides, the authors in \cite{yang2021understanding} studied the AoI in large-scale wireless networks under first-come-first-serve and last-come-first-serve transmission protocols, respectively. Authors in \cite{9085402,9013924} studied the AoI in a large-scale network under a reinforcement learning framework and deep reinforcement learning framework, respectively. In \cite{9360520}, authors considered a large-scale network and the locations of source-destination pairs were Poisson bipolar distributed. Besides, non-preemptive and preemptive queuing disciplines were used to compute the PAoI.  Additionally, in \cite{9467495} authors considered device-to-device communication in a cellular-based IoT network. Specifically, authors computed the network throughput and mean AoI. IoT devices time- and event-triggered traffic based PAoI analysis was provided in \cite{9042825}, while the locations of IoT devices and base stations (BSs) were modeled by two independent Poisson point processes (PPPs).

{\em Analysis of UAV-assisted IoT networks.} Authors in \cite{8570843} optimized the UAV trajectory to minimize the mean PAoI of a source-destination pair networks, in which UAVs behave as communication relays. In \cite{9933782}, authors considered a multipurpose UAV which delivers the data for IoT clusters and packages simultaneously. Authors in \cite{mozaffari2017mobile} investigated the deployment and mobility of multiple UAVs to collect data from IoT devices while minimizing the total transmission power of IoT devices. A UAV-assisted multiple-in-multiple-out (MIMO) communication IoT network was analyzed in \cite{feng2018uav} and the authors jointly designed the transmission duration and power of all devices to maximize the energy efficiency. In \cite{zeng2016throughput}, authors studied the throughput maximization problem in UAV-enabled network, in which the trajectory and transmit power of UAVs are jointly considered. A game theory-based analysis was used in \cite{yan2018game}, in which the authors studied the UAV access selection and BS bandwidth allocation problems while UAVs acting as communication relays between BSs and IoT devices.

Different from the existing literature, we compute the PAoI in a UAV-assisted IoT network, where the locations of IoT devices are modeled by a MCP. Specifically, we investigate the impact of the number of devices and resource allocation techniques on the PAoI.

\subsection{Contribution}
In this paper, our main contribution is computing the mean PAoI in UAV-assisted IoT networks under two load models, (i) bandwidth splitting and (ii) time splitting. We quantify the freshness of the updates in the case of correlated or uncorrelated devices, respectively. The details of the contributions are listed next.

{\em System setup of UAV-assisted IoT networks.} We consider a UAV-assisted IoT network, in which UAVs are deployed above the IoT cluster centers to collect updates from IoT devices. We consider a cluster of devices and they can either monitor one physical process (correlated devices) or several processes (uncorrelated devices). Specifically, we consider that the cluster UAV splits its resources, time or bandwidth, to serve IoT devices to avoid inter-cluster interference.

{\em PAoI analysis under correlated devices.} While the PAoI of uncorrelated devices is widely studied, we mainly study the PAoI of correlated devices. Using tools from stochastic geometry, we model the locations of UAVs and IoT devices by a MCP, which means that IoT devices are uniformly located within the clusters and UAVs are deployed at the cluster centers. Considering that the success probability varies across the devices and is a function of the mean activity probability of devices (interferers) which depends on the success probability of each device, we first provide an equation to compute the mean activity probability of devices and it is solved by iterating.
Besides, we also provide an approximation for computing PAoI under bandwidth splitting, which works well in high LoS environments.

%
%

{\em System-level insights.} By comparing two load models, we show that time splitting works better than bandwidth splitting in UAV-assisted IoT networks for both correlated and uncorrelated devices. Firstly, we observe that time splitting causes higher interference since the channel has a higher probability to be occupied. However, in high LoS scenario, high interference has a limited impact on UAV-assisted networks. This is because: devices are more likely to establish LoS links with cluster UAVs and Non-LoS (NLoS) links with UAVs deployed at other clusters, and the signals received in LoS links are much better than NLoS links. While the UAV splits the time to serve multiple IoT devices, it is more time efficient: UAVs can use the data generation time of a device to serve other devices.

\section{System Model}

\begin{figure}[h]
	\centering
	\includegraphics[width = 1\columnwidth]{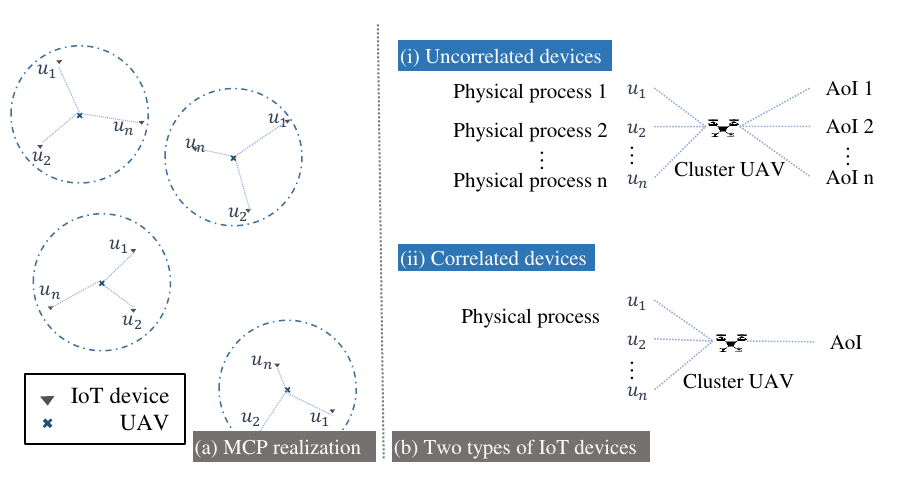}
	\caption{Illustration of system setup: \textbf{(a)} UAVs are deployed above the cluster centers to collect updates from IoT devices and IoT devices are uniformly distributed within the clusters, and \textbf{(b)} (i) uncorrelated devices, which monitor different physical processes, and (ii) correlated devices, which monitor the same physical process.}
	\label{fig_sys_ill}
\end{figure}

An uplink UAV-assisted IoT wireless communication network is considered where IoT devices are spatially clustered and UAVs are deployed above the cluster centers to collect information from IoT devices. The locations of the IoT cluster centers follow a PPP $\Phi_{u}$ with density $\lambda_u$ and IoT devices are uniformly distributed within the clusters with radii $r_c$. Hence, the locations of IoT devices follow a MCP, and UAVs are assumed to hover above the cluster centers at a fixed altitude $h$, as shown in Fig. \ref{fig_sys_ill} (a). The IoT devices send the status updates to their cluster center UAV, and fractional path-loss inversion power control with compensation factor $\epsilon_{\{l,n\}}$ is used in this work, where the subscripts $l,n$ depends on the LoS and NLoS link with the cluster UAV, respectively.
Besides, a standard path-loss model with exponents $\alpha_{\{l,n\}}$ and the Nakagami-m channel fading model are used in this work.  The notations of this work are summarized in Table \ref{notation}.
	\begin{table}\caption{Table of Notations}\label{notation}
	\vspace{-4mm}
	\centering
	\begin{center}
		\resizebox{1\columnwidth}{!}{
			\renewcommand{\arraystretch}{1}
			\begin{tabular}{ {c} | {c}  }
				\hline
				\hline
				\textbf{Notation} & \textbf{Description} \\ \hline
				$\Phi_{u}$, $\lambda_{ u}$&  Locations of IoT devices, IoT clusters density\\\hline
				$\Phi_{i}$ &  Locations of IoT devices that use the same resource block \\\hline
				$r_c$ & MCP disk radius \\\hline
				$h$ & UAV altitude \\\hline
				$ (a,b)$ & LoS/NLoS environment parameters  \\\hline				
				$\rho_{\{l,n\}},p_{u}$ & Power control parameters,  maximum transmission power \\\hline
				 $\sigma^2,\theta$ & Noise power, SINR threshold \\\hline
				$\epsilon_{l},\epsilon_{n}$ & LoS/NLoS path-loss compensation factor \\\hline
				$\alpha_{l},\alpha_{n}$ & LoS/NLoS path-loss exponent \\\hline
				 $m_{l},m_{n}$ & LoS/NLoS fading gain \\\hline
				$\eta_{l},\eta_{n}$ & LoS/NLoS additional loss  \\\hline
				$R_{u,l},R_{u,n}$ & Euclidean transmission distance of LoS/NLoS links \\\hline
				$p_{t,l},p_{t,n}$ & Transmit power of IoT devices of LoS/NLoS links \\\hline
				$p_{r,l},p_{r,n}$ & Received power at the serving UAV of LoS/NLoS link\\\hline
				$G_{l},G_{n}$ & Fading gains of LoS/NLoS links\\\hline
				$N_{d},B$ & Number of IoT devices, available bandwidth of UAVs\\\hline
				$\Delta(t),\bar{\Delta}$ & AoI at time $t$, PAoI\\\hline
				$T_{i},T_{\rm tra},T$ & Inter-arrival time, transmission time of an update, time unit
				\\\hline\hline
		\end{tabular}}
	\end{center}
	\vspace{-4mm}
\end{table} 

Since we consider a cluster of IoT devices, they can either monitor different types of physical random processes or the same type of process, we consider two scenarios for IoT devices in this work: (i) uncorrelated IoT devices, which monitor different types of processes, or (ii) correlated IoT devices, which monitor the same process, as shown in Fig. \ref{fig_sys_ill} (b). To characterize the freshness of the information transmitted by the IoT devices, we compute the mean peak age-of-information (PAoI) at the UAV side for devices in two scenarios. Specifically, we consider two load models: (i) split the bandwidth, and (ii) split the time, to avoid inter-cluster interference. 
To ensure a stable system and avoid infinite PAoI, we consider a non-preemptive discipline which means that IoT devices can only have new updates after the previous updates are successfully transmitted.

 Without loss of generality, we perform the analysis for a typical device, which is randomly selected from a random IoT clusters. Applying Slivnyak's theorem, the center of this cluster (typical cluster), which contains the typical IoT device, is located at the origin. Consequently, both the typical IoT cluster and the UAV are positioned at the origin as well. 

\subsection{Communication Channel}
 As mentioned, we consider a truncated  path-loss inversion power control. Thus, the transmit power of a device depends on established NLoS/LoS channel and is given by
\begin{align}
	p_{t} = \left\{
	\begin{aligned}
		p_{t, l} &=  \rho_l R_{u,l}^{\alpha_{l}\epsilon_{l}}, \text{in the case of LoS},\\
		p_{t, n} &=  \rho_n R_{u,n}^{\alpha_{n}\epsilon_{n}}, \text{in the case of NLoS},\\
	\end{aligned}
	\right.
\end{align}
where the subscript $\{l,n\}$ denotes the established LoS/NLoS channel between the IoT device and the serving UAV: $R_{u,l}$, $\rho_l$, $\epsilon_{l}$ and $\alpha_{ l}$ denote the Euclidean transmission distance, power control parameter, which adjusts the received power at the UAV, compensation factor and path-loss exponent in  LoS scenario, respectively;  $R_{u,n}$, $\rho_n$, $\epsilon_{n}$, $\alpha_{ n}$  denote the same set of parameters for the NLoS scenario. Assuming that the maximum transmit power is $p_u$, the range of $\epsilon_{l,n}$ is
\begin{align}
	0\leq\epsilon_{l,n} \leq  \frac{1}{\alpha_{\{l,n\}}}\log_{\sqrt{r_c^2+h^2}}\bigg(\frac{p_u}{\rho_{\{l,n\}}}\bigg),
\end{align} 
in which $h$ is the altitude of the UAV. Consequently, the received power at the UAV is
\begin{align}
	p_{r}(R_u) = \left\{
	\begin{aligned}
		p_{r, l}(R_{u,l}) &= \eta_{l}p_{t,l} G_{l}R_{u,l}^{-\alpha_{l}}, \text{LoS link},\\
		p_{r, n}(R_{u,n}) &= \eta_{n}p_{t,n} G_{n}R_{u,n}^{-\alpha_{n}}, \text{NLoS link},\\
	\end{aligned}
	\right.
\end{align}
in which  $G_{ l}$ and $G_{ n}$ are the fading gains, which follow Gamma distribution with shape and scale parameters ($m_{ l},\frac{1}{m_{ l}}$) and ($m_{ n},\frac{1}{m_{ n}}$) for LoS and NLoS transmissions, respectively, and $\eta_{ l}$ and $\eta_{ n}$ denote the  mean additional losses in  LoS and NLoS scenario, respectively. Given the horizontal distance $r$ between the serving UAV and the device, the occurrence probability of a LoS link  is given in ~\cite{al2014optimal} as
\begin{align}
	P_{ l}(r) & =  \frac{1}{1+a \exp(-b(\frac{180}{\pi}\arctan(\frac{h}{r})-a))},\label{pl_pn}
\end{align}
where $a$ and $b$ are two environment variables. Consequently, the probability of NLoS link is $P_{ n}(r)=1-P_{  l}(r)$.

\subsection{Load Modeling}
Recall that we consider two scenarios for IoT devices, (i) correlated IoT devices, or (ii) uncorrelated IoT devices, in a cluster, and UAVs fairly allocate their resources to serve these devices to avoid inter-cluster interference.  Let $N_d$ be the number of IoT devices in a cluster and $B$ be the available bandwidth of UAVs. We consider the following two load models and analyze the impact of these two load models (resource block allocation strategy) on the mean PAoI. Let $T$ be the time unit, e.g., sec or min. 

\begin{definition}[Load Model 1: Bandwidth Splitting]
	Equal allocation of bandwidth to serve all the IoT devices. The serving UAV equally splits the available transmission bandwidth among the IoT devices. Thus, each IoT device is assigned a bandwidth equal to $B/N_d$ and the data transmission time of a single device is $N_d T$.
	\end{definition}

\begin{definition}[Load Model 2: Time Splitting]
Exclusive allocation of bandwidth to serve one IoT device. At each transmission time slot, the UAV periodically selects one IoT device and allocates the entire bandwidth to it. Besides, the UAV ensures that all the IoT devices are associated equally to transmit the data in the long term. Thus, each IoT device is assigned a bandwidth equal to $B$ and the data transmission time of a single device is $T$.
\end{definition}

Therefore, in both load models, the interference comes from the devices in other IoT clusters which transmit at the same resource block. Besides, noting that the device only transmits the data when it is active, e.g., has a new update, the interference is also a function of the activity probability of devices. To simplify the system model, we consider the number of devices is the same for all clusters. Therefore, in both Load Model 1 and Load Model 2, the density of IoT devices that might transmit at a specific resource block is $\lambda_u$. Let $\Phi_{i}$ be the point set that contains the locations of IoT devices that transmit at a specific resource block, and $\bar{\pi}$ be the mean activity probability. Therefore, the density of interference is $\bar{\pi}\lambda_{u}$ and $\Phi_i$ is obtained by displacement and thinning of $\Phi_{u}$.

Without loss of generality, we perform the analysis at the typical IoT cluster which is centered at the origin and the typical device which is located within the typical cluster.
 If the signal-to-interference-plus-noise-ratio (SINR) received at the UAV is above a predefined threshold, the updated information from an IoT device is successfully delivered. We define the probability of successful transmission of a typical device to a UAV link as success probability. The conditional success probability, which is conditioned on an arbitrary but fixed realization of $\Phi_{u}$ and $\Phi_i$, is provided next.

\begin{definition}[Conditional Success Probability]
The conditional success probability is defined as,
\begin{align}
	\label{eq_Ps}
	P_s(R_u) = \mathbb{P}({\rm SINR}(R_u)>\theta|\Phi_{i},\Phi_{u}),
\end{align}
in which $\theta$ is the SINR threshold,
\begin{align}
	{\rm SINR}(R_u) =& \frac{p_{r}(R_u)}{I+\sigma^2},
\end{align}
where $I$ denotes the aggregated interference,
\begin{align}
	I =& \sum_{u_i\in\Phi_{i}^{ll}}\eta_{ l} G_l p_{t,l,u_i}D_{u_i}^{-\alpha_{l}}+\sum_{u_i\in\Phi_{i}^{nl}}\eta_{l} G_l p_{t,n,u_i}D_{u_i}^{-\alpha_{l}}\nonumber\\
	&+\sum_{u_i\in\Phi_{i}^{ln}}\eta_{ n} G_n p_{t,l,u_i}D_{u_i}^{-\alpha_{n}}\nonumber\\
	&+\sum_{u_i\in\Phi_{i}^{nn}}\eta_{n} G_n p_{t,n,u_i}D_{u_i}^{-\alpha_{n}},\label{eq_I}
\end{align}
in which $i_0$ denotes the location of the typical device, $\Phi_{i}^{ll}$, $\Phi_{i}^{ln}$, $\Phi_{i}^{nl}$ and $\Phi_{i}^{nn}$ are subsets of $\Phi_{i}\setminus \{i_o\}$ denote the locations of interfering IoT devices which establish LoS/NLoS links with their cluster UAVs and have LoS/NLoS links with the typical UAV, respectively, $D_{\{\cdot\}}$ denotes the distances between the interfering users and the typical UAV, and $p_{t,\{l,n\},u_i}$ denote the transmission power of interferer, which establish LoS/NLoS links with their cluster UAVs, respectively.
\end{definition} 

Since we consider the cluster UAV serves multiple IoT devices and the locations of devices are randomly distributed within the IoT clusters, the success probabilities vary with devices.  Therefore, we define the SINR meta distribution which computes the complementary cumulative distribution function (CCDF) of success probability \cite{haenggi2015meta,haenggi2021meta1,haenggi2021meta2}. 

 \begin{definition}[SINR Meta Distribution]
The SINR meta distribution, defined as the CCDF of the conditional success probability, is defined in \cite{haenggi2015meta} and given by
\begin{align}
	\bar{F}_{P_s}(\gamma) &= \mathbb{P}(P_s(R_u) > \gamma),
\end{align} 
where $\gamma \in [0,1]$.
\end{definition} 

\subsection{Age of Information}

As mentioned, AoI characterizes the freshness of the data. Let $\Delta(t)$ track the AoI evolution at time slot $t$. Let $k$ be the index of the update at IoT devices and $G(k)$ denote the generation time of the $k$-th update. Therefore, AoI evolution of the $k$-th update is given in \cite{kaul2012real}, computed by
\begin{align}
	\label{eq_AoI}
	\Delta_{k}(t+1) &=\left\{ 
	\begin{aligned}
		\Delta_{k}(t)+1,&\quad \text{if transmission fails},\\
	    t+1-G(k), &\quad  \text{otherwise}.\\
	\end{aligned} \right.
\end{align}
At the side of the serving UAV, $\Delta(t)$ increases with time and drops upon the arrival of new updates. In this work, we consider the update generation and the successful transmission occurs at the end of the time slot. Besides, for a unit time slot $T$, the generation probability of an update is $\lambda_a$ and the generation processes of IoT devices are independent and follow a geometric distribution.

\begin{figure}[h]
	\centering
	\includegraphics[width = 0.7\columnwidth]{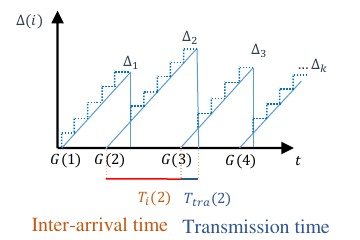}
	\caption{Illustration of the AoI evolution of a typical physical process of uncorrelated devices.}
	\label{fig_aoi_ill}
\end{figure}

In this work, we characterize the mean peak AoI, which is defined as the average value of age resulting immediately before the receiving of a new update, as shown in Fig. \ref{fig_aoi_ill}.
\begin{definition}[Mean PAoI] The mean PAoI is defined as the mean value of PAoI, in which PAoI is measured immediately before the reception of an update from UAV side,
\begin{align}
\overline{\Delta} = \mathbb{E}[\Delta_{k}|\Phi_{u},\Phi_{i}] = \mathbb{E}[T_{i}+T_{\rm tra}|\Phi_{u},\Phi_{i}],
\end{align}
in which $\Delta_{k}$ denotes the PAoI of the $k$-th updates, $T_{i}$ is the inter-arrival time and $T_{\rm tra}$ denotes the transmission time and the expectation sign is over the locations of the devices and the fading gains.
\end{definition}

In the case of uncorrelated devices, AoI works independently for different updates of  physical processes, and the evolution process is shown in Fig. \ref{fig_aoi_ill}.

\begin{figure}[h]
	\centering
	\includegraphics[width = 1\columnwidth]{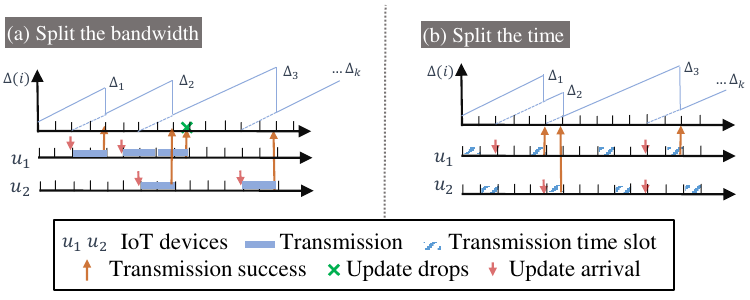}
	\caption{Illustration of the AoI evolution of  correlated devices, \textbf{(a)} Load Model 1, split the bandwidth, and \textbf{(b)} Load Model 2, split the time.}
	\label{fig_aoi_ill2}
\end{figure}

In the case of correlated devices, we consider all the devices working independently and assume that the cluster UAV only keeps the most recent updates. It means that AoI only drops when the arrival update is generated later than the current information UAV kept, as shown in Fig. \ref{fig_aoi_ill2} (a), where the green crosses denote that the UAV drops the updates. Besides, we plot the AoI evolution under two load models in Fig. \ref{fig_aoi_ill2} where the downward arrows denote the generation of new updates and the upward arrows denote the successful transmission of the update. Moreover, Fig. \ref{fig_aoi_ill2} (a) shows that the devices can transmit the update immediately but have a longer transmission time, and Fig. \ref{fig_aoi_ill2} (b) shows that the devices need to wait for the transmission time slot but have a shorter transmission time.

The main difference between the PAoI of correlated and uncorrelated devices is that the PAoI of correlated only contains one successful transmission while the PAoI of uncorrelated contains two successful transmission since we assume that devices can only have new update when the previous date is successfully transmitted.

In the following sections, we analyze the mean PAoI under two load models and show the impact of the number of devices on the system performance.
\section{Success Probability Analysis}
In this section, we provide the analysis of success probability under two load models and two scenarios for IoT devices. To do so, we first compute the probability density function (PDF) of the transmission power of devices and then compute the success probability. Since the conditional success probability is a function of the density of interferers, we provide the equations of the conditional activity probability of devices in this section, but the details of conditional activity probabilities are provided in the next section.

As mentioned previously, we consider a truncated inverse power control in this uplink network. The transmit power of devices is a function of transmission distance and the distribution of transmit power of IoT devices is given in the following lemma.
\begin{lemma}[Distribution of the Transmit Power] The PDF of the transmit power of a general active device in the case of establishing a LoS/NLoS with cluster UAV is
\begin{align}
&f_{p_{t,l}}(p) \nonumber\\
&=\left\{
\begin{aligned}
	\frac{2P_l((\frac{p}{\rho_l})^{1/(\epsilon_l\alpha_l)})(\frac{p}{\rho_l})^{2/(\epsilon_l\alpha_l)-1}}{(\epsilon_l\alpha_l)\rho_l r_c^2}&, \rho_l <p < p_{u},\\
	\int_{p_u}^{\infty}\frac{2P_l((\frac{x}{\rho_l})^{1/(\epsilon_l\alpha_l)})(\frac{x}{\rho_l})^{2/(\epsilon_l\alpha_l)-1}}{(\epsilon_l\alpha_l)\rho_l r_c^2}{\rm d}x	&,  p = p_{u},\\
\end{aligned}
\right.\\
&f_{p_{t,n}}(p) \nonumber\\ &=\left\{
\begin{aligned}
	\frac{P_n((\frac{p}{\rho_n})^{1/(\epsilon_n\alpha_n)})p^{2/(\epsilon_n\alpha_n)-1}}{(\epsilon_n\alpha_n)\rho_n r_c^2}&, \rho_n <p < p_{u},\\
	\int_{p_u}^{\infty}\frac{2P_n((\frac{x}{\rho_n})^{1/(\epsilon_n\alpha_n)})(\frac{x}{\rho_n})^{2/(\epsilon_n\alpha_n)-1}}{(\epsilon_n\alpha_n)\rho_n r_c^2}{\rm d}x	&,  p = p_{u},\\
\end{aligned}
\right.
\end{align}
in which $P_l(x)$ and $P_n(x)$ are the LoS/NLoS probabilities as defined in (\ref{pl_pn}).
\end{lemma}
\begin{IEEEproof}
Similar to \cite{6786498,7894264}, in the uplink transmission of a MCP, the transmission distance has the distribution $f_{R_{u,l}}(r) = \frac{2\int_{0}^{x}P_{l}(x){\rm d}x}{r_{c}^2}$, and the transmit power of a device is $p_{t,l} = \rho_l R_{u,l}^{\alpha_l\epsilon_l}$. The PDF of the transmit power is obtained by inserting $r = (p_{t,l}/\rho_l)^{1/(\epsilon_l\alpha_l)}$ into $f_{R,l}(r)$.
\end{IEEEproof}

 Before analyzing the interference, we first need to obtain the activity probability of the devices.  Recall that we assume a non-preemptive discipline, devices can only have new updates when the previous update is successfully transmitted and the device is active when it has a new update.

\begin{figure}[ht]
	\centering
	\includegraphics[width = 0.8\columnwidth]{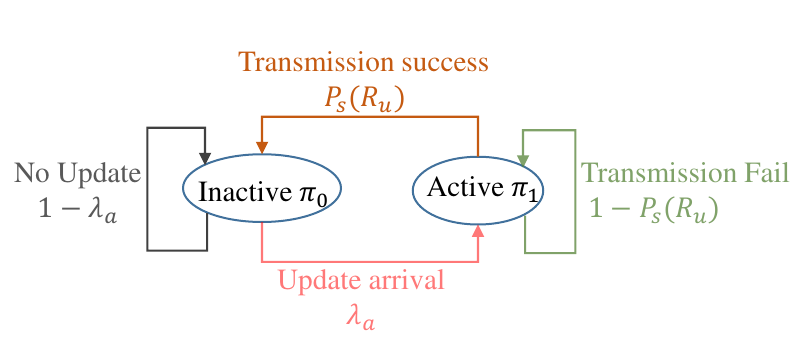}
	\caption{Illustration of the state transfer under Load Model 1 or $N_d = 1$.}
	\label{fig_ill_stead}
\end{figure}

 In the case of Load model 1, the state transmission diagram is shown in Fig. \ref{fig_ill_stead}. The steady state probability for both load models when $N_d=1$ is
 \begin{align}
 \pi_1 = \frac{\lambda_{a}}{\lambda_{a}+P_s(R_u)}.
 \end{align}
The activity probabilities when $N_d>1$ are provided in the following lemma. The activity probability of Load Model 2 is slightly different owing to the waiting of the transmission time slot, hence, in the following lemma, we provide the formulas first and details are provided in the PAoI analysis under each load model in the next section. 

\begin{lemma}[Conditional Activity Probability]
Conditioned on the locations of devices, the activity probability of the typical device under Load Model 1 when $N_d>1$ is
\begin{align}
	\label{eq_condition_activeprob1}
	\pi_{1,l_1} = \frac{\lambda_{a}^{'}}{\lambda_{a}^{'}+P_s(R_u)},
\end{align}
in which $P_s(R_u)$ is the success probability of the device as defined in (\ref{eq_Ps}), and $\lambda_{a}^{'}$ reflects the impact of the number of devices on the conditional activity probability, which will be discussed in  Section \ref{section_meanPAoI}. 	

Conditioned on the locations of devices, the activity probability of the typical device under Load Model 2 is
\begin{align}
	\label{eq_condition_activeprob2}
	\pi_{1,l_2}  = P_s(R_u)\lambda_{a}^{''}+(1-P_s(R_u)),
\end{align}
in which $\lambda_{a}^{''}$ reflects the impact of $N_d$ on the conditional activity probability, 
more details are provided in Section \ref{section_meanPAoI}.
\end{lemma}

It is clear that the conditional activity probability is a function of the success probability, which depends on the realization of $\Phi_{u}$, $\Phi_{i}$, and the interference density, which depends on the mean activity probability. To solve this problem, we propose the following method which can compute the mean activity probability accurately but requires an iteration over the arrival rate. To do so, we first let $\bar{\pi}$ be the mean activity probability, $\bar{\pi}=\mathbb{E}[\{\pi_{1,l_1},\pi_{1,l_2}\}]$, and compute the first and the second moments of the conditional success probability given $\bar{\pi}$. Therefore, we first derive the conditional probability and the first and the second moments.

\begin{lemma}[Conditional Success Probability]
Conditioned on the transmission distance between the IoT device and the cluster UAV, the success probability in LoS/NLoS scenario is given by
\begin{align}
	 &P_{s,l}(R_u)  
= P_l(R_u)\sum_{k = 1}^{m_l}\binom{m_l}{k}(-1)^{k+1}\exp(-g_l(R_u)\sigma^2)\nonumber\\
	& \prod_{c_1,c_2\in\{l,n\}}\exp\bigg(-2\pi\bar{\pi}\lambda_u\int_{h}^{\infty}\int_{0}^{p_u}\bigg[1-\kappa(g_l(R_u),c_1,c_2)\bigg]\nonumber\\
	&\times zP_{c_1}(\sqrt{z^2-h^2})f_{p_{t,c_2}}(p){\rm d}p{\rm d}z\bigg),\label{eq_ps_l}\\
	&P_{s,n}(R_u)  
	= P_n(R_u)\sum_{k = 1}^{m_n}\binom{m_n}{k}(-1)^{k+1}\exp(-g_n(R_u)\sigma^2)\nonumber\\
	& \prod_{c_1,c_2\in\{l,n\}}\exp\bigg(-2\pi\bar{\pi}\lambda_u\int_{h}^{\infty}\int_{0}^{p_u}\bigg[1-\kappa(g_n(R_u),c_1,c_2)\bigg]\nonumber\\
	&\times zP_{c_1}(\sqrt{z^2-h^2})f_{p_{t,c_2}}(p){\rm d}p{\rm d}z\bigg),\label{eq_ps_n}
\end{align}
where $g_l(r) = k\beta_2(m_l)m_l\theta r^{(1-\epsilon_{l})\alpha_{ l}}(\rho_l\eta_{ l})^{-1}$, $g_n(r) = k\beta_2(m_n)m_n\theta r^{(1-\epsilon_{n})\alpha_{n}}(\rho_n\eta_{n})^{-1}$, $\beta_2(m) = (m!)^{-1/m}$ and  $\bar{\pi}$ is the mean activity probability,
\begin{align}
\kappa(g(r),c_1,c_2) = \bigg(\frac{m_{c_1}}{m_{c_1}+g(r)\eta_{c_2}p z^{-\alpha_{c_2}}}\bigg)^{m_{c_1}},
\end{align}
\end{lemma}
\begin{IEEEproof}
Similar to \cite[Theorem 1]{Mbref}.
\end{IEEEproof}

Consequently, the first and the second moments of the conditional success probability are given in the following lemma. The moment of the conditional success probability is a function of $\bar{\pi}$ since the mean activity probability influences the density of the interferers.

\begin{lemma}[The Moments of The Conditional Success Probability]\label{lemma_Mb}
The moment of the conditional success probability is derived	by taking the expectation over the locations of interferers and distance to the serving UAV,	given by
\begin{align}
	M_b(\bar{\pi}) &= 	M_{b,l}(\bar{\pi})+ M_{b,n}(\bar{\pi})=\mathbb{E}[P_{s, l}^{b}(R_u)]+\mathbb{E}[P_{s, n}^{b}(R_u)],
\end{align}
where $b = 1,2$ denotes the first and the second moment, and are given by
\begin{align}
	M_{b,l}(\bar{\pi}) =&  \sum_{k_1 = 1}^{m_l}\cdots\sum_{k_b = 1}^{m_l}\binom{m_l}{k}\cdots\binom{m_l}{k_b}(-1)^{k_1\cdots k_b+b}\nonumber\\
	&\int_{0}^{r_c}P_l(r)\mathcal{L}(g_{l,1}(r),\cdots,g_{l,b}(r))f_{ R_u}(r){\rm d}r,\nonumber\\
	M_{b,n}(\bar{\pi}) =& \sum_{k_1 = 1}^{m_n}\cdots\sum_{k_b = 1}^{m_n}\binom{m_n}{k}\cdots\binom{m_n}{k_b}(-1)^{k_1\cdots k_b+b}\nonumber\\
	&\int_{0}^{r_c}P_n(r)\mathcal{L}(g_{n,1}(r),\cdots,g_{n,b}(r))f_{ R_u}(r){\rm d}r,\label{eq_Mb}
\end{align}
where $g_{l,i}(r) = k_i\beta_2(m_l)m_l\theta r^{(1-\epsilon_{l})\alpha_{ l}}(\rho_l\eta_{ l})^{-1}$,  $g_{n,i}(r) = k_i\beta_2(m_n)m_n\theta r^{(1-\epsilon_{n})\alpha_{ n}}(\rho_n\eta_{ n})^{-1}$, and $\mathcal{L}(g_{1}(r),\cdots,g_{b}(r))$ is shown in Appendix \ref{app_laplace}.
\end{lemma}
\begin{IEEEproof}
	Similar to \cite[Lemma 2]{Upref}.
\end{IEEEproof}


To simplify the analysis of this work, we use the beta approximation to approximate the meta distribution, and our previous work \cite[Lemma 2]{Upref} and existing literature \cite{haenggi2015meta} haven shown the accuracy of the beta approximation. Therefore, the SINR meta distribution of the proposed uplink network is give in the following lemma.
\begin{lemma}[SINR Meta Distribution] The SINR meta distribution, which is defined as the CCDF of the conditional success probability is approximated by
\begin{align}
\bar{F}_{P_s}(\gamma) &\approx1-I_{\gamma}(\frac{M_1(\bar{\pi})(M_1(\bar{\pi})-M_2(\bar{\pi}))}{M_2(\bar{\pi})-M_1^2(\bar{\pi})},\nonumber\\
&\frac{(M_1(\bar{\pi})-M_2(\bar{\pi}))(1-M_1(\bar{\pi}))}{M_2(\bar{\pi})-M_1^2(\bar{\pi})}),
\end{align}
where $M_1(\bar{\pi})$ and $M_2(\bar{\pi})$ are the first and the second moment of the proposed network,
\begin{align}
	I_x(a,b) &= \frac{\int_{0}^{x}t^{a-1}(1-t)^{b-1}{\rm d}t}{B(a,b)},\nonumber\\
	B(a,b) &= \int_{0}^{1}t^{a-1}(1-t)^{b-1}{\rm d}t.
\end{align}
\end{lemma}

Note from (\ref{eq_condition_activeprob1}) and (\ref{eq_condition_activeprob2}) that, the conditional activity probability is a function of success probability. Thus, we can obtain the distribution of conditional probability as 
\begin{align}
\mathbb{P}(	\pi_{1,l_1} < x)  &= \mathbb{P}\bigg(\frac{\lambda_{a}^{'}}{\lambda_{a}^{'}+P_{s}(R_u)}<x\bigg)\nonumber\\
 &= \mathbb{P}\bigg(P_{s}(R_u) > \frac{1-x}{x}\lambda_{a}^{'}\bigg) = \bar{F}_{P_{s}}\bigg(\frac{1-x}{x}\lambda_{a}^{'}\bigg),\nonumber\\
 \mathbb{P}(	\pi_{1,l_2} < x)  &= \mathbb{P}(P_s(R_u)\lambda_{a}^{''}+(1-P_s(R_u))<x)
\nonumber\\
&= \bar{F}_{P_{s}}\bigg(\frac{1-x}{1-\lambda_{a}^{''}}\bigg),
\end{align}
in which $\lambda_{a}^{'}$ and $\lambda_{a}^{''}$ are provided in (\ref{eq_lambda_a_p}) and (\ref{eq_lambda_a_pp}) which reflects the influence of the number of serving devices.
 Therefore, the mean activity probability is obtained by taking the expectation, as shown in the following theorem.

\begin{theorem}[Activity Probability] 
\label{theorem_activeprob}
The unconditional activity probability $\hat{\pi}$ is given by
\begin{align}
	\label{eq_mean_pi}
	\bar{\pi}  =& \int_{0}^{1}\bar{F}_{P_s}(g(x)){\rm d}x \nonumber\\
	=&\int_{0}^{1}I_{g(x)}(\frac{M_1(\bar{\pi})(M_1(\bar{\pi})-M_2(\bar{\pi}))}{M_2(\bar{\pi})-M_1^2(\bar{\pi})},\nonumber\\
	&\frac{(M_1(\bar{\pi})-M_2(\bar{\pi}))(1-M_1(\bar{\pi}))}{M_2(\bar{\pi})-M_1^2(\bar{\pi})}){\rm d}x,
\end{align}
in which $g(x)$ is $\min(1,(\frac{1-x}{x}\lambda_{a}^{'})^{+})$ in Load Model 1 and is $\min(1,(\frac{1-x}{1-\lambda_{a}^{''}})^{+})$ in Load Model 2, where $(x)^{+} = \max(x,0)$.
Observe that both sides of (\ref{eq_mean_pi}) contain the unknown variable $\bar{\pi}$, e.g., $M_1(\bar{\pi})$ and $M_2(\bar{\pi})$ are functions of $\bar{\pi}$, this equation can be solved by iterations: for a specific $\bar{\pi}$, we find a typical $\lambda_{a}$ which let the equal sign hold.
\end{theorem}

After obtaining the mean activity probability and conditional success probability of an individual device, we are able to proceed to the next step: compute the mean PAoI.

\section{Mean PAoI Analysis}
\label{section_meanPAoI}
In this section, we provide the analysis to compute the mean PAoI under two load models, two scenarios for IoT devices (correlated and uncorrelated), respectively. Specifically, we consider an approximation that all the correlated devices have the same success probability. Besides, while the analysis results of mean PAoI under $N_d>1$ are provided in the following two subsections, the results of $N_d=1$ are the same for both load models, which is
\begin{align}
\bar{\Delta} & = \mathbb{E}_{R_u}[T_i+T_{\rm tra}] = \mathbb{E}_{R_u}\bigg[\frac{2T}{P_s(R_u)}+\frac{T}{\lambda_{a}}\bigg],
\end{align}
in which $\frac{T}{P_s(R_u)}+\frac{T}{\lambda_{a}}$ computes the mean inter-arrival time and $\frac{T}{P_s(R_u)}$ denotes the mean transmission time.

\subsection{Mean PAoI Under Load Model 1}
In this subsection, we derive the mean PAoI under Load Model 1, in which the cluster UAV splits the bandwidth to serve correlated IoT devices.
Recall that $N_d$ denotes the number of devices within the cluster, and for a unit of time $T$, the arrival rate of an update is $\lambda_a$ and the generation process follows a geometric distribution. When the UAV splits the bandwidth to serve devices, the transmission time increases with the increasing of the number of  serving devices $T_{m1} = N_d T$. The device is active when it has an update to transmit or when it is transmitting updates. Hence, the activity probability, which computes as a time average ratio, in Load Model 1 is
\begin{align}
\pi_1 = \frac{\bar{T}_{\rm tra}}{\bar{T}_{\rm tra}+\bar{T}_i},
\end{align}
in which $\bar{T}_{\rm tra}$ and $\bar{T}_i$ are mean transmission time and mean data inter-arrival time, respectively. The distribution of successful transmission time follows a geometric distribution 
\begin{align}
	\mathbb{P}(T_{\rm tra} = n T_{m1} ) &= (1-P_s(R_u))^{n-1}P_s(R_u), n = 1,2,3,\cdots, \nonumber\\
	\bar{T}_i &= \frac{1}{\lambda_a}.
\end{align}
Consequently, the conditional activity probability and $\lambda_{a}^{'}$ in (\ref{eq_condition_activeprob1}) are, respectively, given by
\begin{align}
\label{eq_lambda_a_p}
\pi_{1,l_1} &= \frac{N_d\lambda_a }{N_d\lambda_a+P_s(R_u)}, \nonumber\\
\lambda_{a}^{'} &= N_d\lambda_a.
\end{align}

\subsubsection{Correlated Devices}

Recall that we randomly choose one device from the cluster as the typical device and use the notation $\Delta_{i,k}(t)$ to denote the AoI of the $i$-th device and $\Delta_{i,k}$ to denote the PAoI of the $i$-th device, in which the subscript $k$ denotes the index of the update. 
For the typical device $o$, the probability mass function (PMF) of $\Delta_{o}$ is given in the following lemma.

\begin{lemma}[Distribution of the PAoI under Load Model 1]\label{lemma_PMF_PAOI_L1}
The PMF of the PAoI under Load Model 1 is given by,
\begin{align}
	&f_{\Delta_{o}}(n\times T) = \mathbb{P}(\Delta_{o,k} = n) \nonumber\\
	&= \sum_{n_1,n_2\in\mathcal{N}_1}\lambda_{a}P_{s}(R_u)(1-\lambda_{a})^{n_1-1}(1-P_{s}(R_u))^{n_2-1},\nonumber\\
	& (n = N_d+2,N_d+3,\cdots),\nonumber\\
	&f_{\Delta_{o}}(1\times T)	=f_{\Delta_{o}}(2\times T)	=\cdots = f_{\Delta_{o}}((N_d+1)\times T) =0,
\end{align}
where $\mathcal{N}_1=\{n_1,n_2:n_1+n_2 N_d = n\}$. Accordingly, the CCDF of PAoI is given by
\begin{align}
&	\bar{F}_{\Delta_{o}}(n\times T) = \mathbb{P}(\Delta_{o,k} > n) \nonumber\\
	&= \sum_{l = n+1}^{\infty}\sum_{n_1+n_2 N_d = l}\lambda_{a}P_{s}(R_u)(1-\lambda_{a})^{n_1-1}(1-P_{s}(R_u))^{n_2-1},\nonumber\\
	&( n = N_d+1,N_d+2,\cdots),\nonumber\\
&	\bar{F}_{\Delta_{o}}(1\times T) =\bar{F}_{\Delta_{o}}(2\times T) = \cdots =\bar{F}_{\Delta_{o}}((N_d+1)\times T) =1.
\end{align}
\end{lemma}
\begin{IEEEproof}
The PAoI of the typical device in the case of Load Model 1 is computed by
\begin{align}
&f_{\Delta_{o}}(n\times T) = \mathbb{P}(\Delta_{o,k} = n) \nonumber\\
&= \mathbb{P}(T_i = n_1,T_{\rm tra} = n-n_1)= \mathbb{P}(T_i = n_1)\mathbb{P}(T_{\rm tra} = n-n_1),
\end{align}
proof completes by inserting the PMFs of inter-arrival time and success transmissions time, respectively, and both are geometric distributed.

Different from the PAoI of $N_d = 1$, which requires two successful transmission, PAoI of multiple devices only requires one. This is because the update can be generated when other updates are currently transmitting.
\end{IEEEproof}

Note that the reason the value of PAoI starts from $N_d+2$ is that we assume the update generates and the data transmit at the end of the time slot. Hence, the minimum value of PAoI happens when the previous data is successfully transmitted and the update data generates immediately, which takes one $T$, and successfully transmits immediately, which takes $N_d T$,  after the transmission of the previous data.

Obtaining the PMF $\Delta_{o}$ of an individual device, we are able to derive the mean PAoI of correlated devices, which is given in the following theorems.
\begin{theorem}[Mean PAoI under Load Model 1 and Correlated Devices]
\label{theorem_mPAoI_l1}
 In the case of correlated devices, the mean PAoI under the Load Model 1 is given by
\begin{align}
		\label{eq_MeanPAoI_l1}
	\bar{\Delta}_{l_1} &\approx \mathbb{E}_{R_u}[\bar{F}_{\Delta_{o}}^{N_d}(n\times T)],
\end{align}	
in which the approximation sign comes from the fact that we assume all the devices have the same success probability.
\end{theorem}
\begin{IEEEproof}
As mentioned, we approximate that these devices have the same success probability.	Conditioned on $R_u$, the CCDF of the PAoI in the case of Load Model 1 and correlated devices is computed by
	\begin{align}
&\bar{F}_{\Delta_{l_1}}(n\times T) = \mathbb{P}(\Delta_{1}>n,\cdots,\Delta_{N_d}>n)\nonumber\\
&=\mathbb{P}(\Delta_{1}>n)\cdots\mathbb{P}(\Delta_{N_d}>n) \nonumber\\
&= \bar{F}_{\Delta_{1}}(n\times T)\cdots\bar{F}_{\Delta_{N_d}}(n\times T)\nonumber\\
&= \bar{F}_{\Delta_{o}}^{N_d}(n\times T),
	\end{align}
	proof completes by taking the expectation over $R_u$.
\end{IEEEproof}

Besides, while the number of devices increases, the data arrival rate, and success transmission probability from the perspective of the serving UAV increases. In the following remark, we propose an approximation that considers a modified arrival rate and a modified success probability.

\begin{remark}[Approximation of Mean PAoI under Load Model 1 and Correlated Devices]
\label{remark_app}
At high values of conditional success probability (high LoS probability areas), the mean PAoI under Load model 1 can be computed by
\begin{align}
\bar{\Delta}_{l_1} \approx&	\mathbb{E}_{R_u}\bigg[\frac{ T_{m1}}{\hat{P}_{s}(R_u)}+\frac{T}{\hat{\lambda}_{a}}\bigg]\nonumber\\
	=& \int_{0}^{r_c}P_l(r)\bigg(\frac{T_{m1}}{\hat{P}_{s,l}(r)}+\frac{T}{\hat{\lambda}_{a}}\bigg)f_{R_u}(r){\rm d}r\nonumber\\
	&+\int_{0}^{r_c}P_n(r)\bigg(\frac{T_{m1}}{\hat{P}_{s,n}(r)}+\frac{T}{\hat{\lambda}_{a}}\bigg)f_{R_u}(r){\rm d}r,
\end{align}
in which the approximation sign comes from the fact that we use the same success probability of all the devices, and
\begin{align}
	\hat{\lambda}_{a} &= (1-(1-\lambda_a)^{N_d}),\nonumber\\
\hat{P}_{s}(R_u) &= (1-(1-P_s(R_u))^{N_d}).
\end{align}
However, this approximation shows gaps at low LoS probability areas and the mean PAoI behaves more close to a single device in such environment. More details are discussed in Section \ref{sect_numerical}.
\end{remark}

\subsubsection{Uncorrelated Devices}
In the case of uncorrelated devices, we compute the mean PAoI directly, which is composed of mean data transmission time and mean data inter-arrival time.
\begin{theorem}[Mean PAoI under Load Model 1 and Uncorrelated Devices] In the case of uncorrelated devices, the mean PAoI under the Load Model 1 is given by
\begin{align}
	\bar{\Delta}_{l_1}  =& \mathbb{E}_{R_u}\bigg[\frac{2\times T_{m1}}{P_s(R_u)}+\frac{T}{\lambda_{a}}\bigg]\nonumber\\
	=& \int_{0}^{r_c}P_l(r)\bigg(\frac{2\times T_{m1}}{P_{s,l}(r)}+\frac{T}{\lambda_{a}}\bigg)f_{R_u}(r){\rm d}r\nonumber\\
	&+\int_{0}^{r_c}P_n(r)\bigg(\frac{2\times T_{m1}}{P_{s,n}(r)}+\frac{T}{\lambda_{a}}\bigg)f_{R_u}(r){\rm d}r,
\end{align}
	in which $P_{s,\{l,n\}}(R_u)$ are provided in (\ref{eq_ps_l}) and (\ref{eq_ps_n}).
\end{theorem}

\subsection{Mean PAoI Under Load Modeling 2}

In this subsection, we derive the mean PAoI under Load Model 2, in which the cluster UAV splits the time and serves the IoT devices periodically. Since the UAV serves the devices periodically with the period  $T_{m2} = N_d T$ the PMF of the successful transmission time is
\begin{align}
	\mathbb{P}(T_{\rm tra} = (n-1) \times T_{m2}+T) &= (1-P_s(R_u))^{n-1}P_s(R_u),\nonumber\\
	&( n = 1,2,3,\cdots).
\end{align}
 From the perspective of the cluster UAV, the transmission time slot is occupation if the serving device has an update or the previous transmission fails, therefore, the conditional transmission time slot occupied probability is
\begin{align}
	\label{eq_lambda_a_pp}
	\pi_{1,l_2}  &= P_s(R_u)\lambda_{a}^{''}+(1-P_s(R_u)), \nonumber\\
	\lambda_{a}^{''} &= 1-(1-\lambda_a)^{N_d},
\end{align}
in which $P_s(R_u)\lambda_{a}^{''}$ denotes the arrival probability of new update and $(1-P_s(R_u))$ denotes the probability of re-transmission.

\subsubsection{Correlated Devices}
\begin{figure}[ht]
	\centering
	\includegraphics[width = 1\columnwidth]{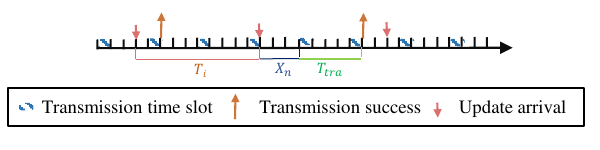}
	\caption{Illustration of the PAoI of Load Model 2.}
	\label{fig_ill_lm2_waiting}
\end{figure}

Different from Load Model 1, in Load Model 2, the device can only transmit the update during the serving time slot. That is, the new update needs to wait. Let $X_n$ be the average waiting time of the updates, and $\Delta_{o}^{'} = T_{i}+T_{\rm tra}$ be the AoI without waiting, as shown in Fig. \ref{fig_ill_lm2_waiting}. The PMF of $\Delta_{o}^{'}$ is shown in the following lemma.

\begin{lemma}[Distribution of $\Delta_{o}^{'}$ under Load Model 2]
	The PMF of $\Delta_{o}^{'}$ under Load Model 2 is given by,
	\begin{align}
		f_{\Delta_{o}^{'}}(n\times T) &= \mathbb{P}(\Delta_{o,k}^{'} = n\times T) \nonumber\\
		&= \sum_{n_1 =0}^{\lfloor\frac{n-2}{N_d}\rfloor}\lambda_{a}P_{s}(R_u)(1-\lambda_{a})^{n-2-N_d n_1}(1-P_{s}(R_u))^{n_1},\nonumber\\
		& (n = 3,4,\cdots),\nonumber\\
		f_{\Delta_{o}^{'}}(1\times T)	& = f_{\Delta_{o}^{'}}(2\times T) =0,
	\end{align}
	accordingly, the CCDF of $\Delta_{o}^{'}$ is given by
	\begin{align}
		&\bar{F}_{\Delta_{o}^{'}}(n\times T) = \mathbb{P}(\Delta_{o,k} > n\times T) \nonumber\\
		&= \sum_{l = n+1}^{\infty}\sum_{n_1 =0}^{\lfloor\frac{n-2}{N_d}\rfloor}\lambda_{a}P_{s}(R_u)(1-\lambda_{a})^{n-2-N_d n_1}(1-P_{s}(R_u))^{n_1},\nonumber\\
		& (n = 3,4,\cdots),\nonumber\\
	&	\bar{F}_{\Delta_{o}^{'}}(1\times T)	 = \bar{F}_{\Delta_{o}^{'}}(2\times T) =1.
	\end{align}
\end{lemma}
\begin{IEEEproof}
	Similar to the proof of Lemma \ref{lemma_PMF_PAOI_L1}, just note that the minimum value of $\Delta_{o}^{'}$ is $3T$ since we ignore the waiting time, and it happens when the update successfully transmits immediately after it generates.
\end{IEEEproof}

After obtaining the PMF and CCDF of the PAoI of an individual device, we are able to derive the mean PAoI of correlated devices, which is given in the following theorem.

\begin{theorem}[Mean PAoI under Load Model 2 and Correlated Devices]  In the case of correlated devices, the mean PAoI under the Load Model 2 is given by
\begin{align}
	\label{eq_MeanPAoI_l2}
	\bar{\Delta}_{l_2} &\approx \mathbb{E}_{R_u}[\bar{F}_{\Delta_{o}^{'}}^{N_d}(n\times T)] + X_n \times T,
\end{align}
in which the approximation sign comes from the fact that we assume all the devices have the same success probability, and $ X_n$ characterizes the average waiting time, the number of the time slots between the data arrival and the transmission time slot,
\begin{align}
	X_n &= \sum_{k = 1}^{N_d-1}x_n(k)(N_d-k-1)+x_n(N_d)(N_d-1),\nonumber\\
	x_n(k) &= (1-\lambda_a)^{k-1}\frac{\lambda_{a}}{1-(1-\lambda_a)^{N_d}}.
\end{align}
\end{theorem}
\begin{IEEEproof}
	Similar to the proof of Theorem \ref{theorem_mPAoI_l1} and $X_n$ is computed by
\begin{align}
&	\mathbb{P}(T_w = (N_d-k-1)\times T) = \mathbb{P}(T_a = k\times T)\nonumber\\
&\quad+\mathbb{P}(T_a = (k+N_d)\times T)\nonumber\\
	&\quad+\mathbb{P}(T_a = (k+2N_d)\times T)+\cdots \nonumber\\
	&= \sum_{i = 0}^{\infty} \mathbb{P}(T_a = (k+iN_d)\times T) = \sum_{i = 0}^{\infty} (1-\lambda_{a})^{k+iN_d-1}\lambda_{a},\nonumber\\
	X_n &= \sum_{k = 1}^{N_d-1}T_w\mathbb{P}(T_w = (N_d-k-1)\times T)\nonumber\\
	&\quad+(N_d-1)\mathbb{P}(T_w = N_d\times T),
\end{align}
in which $T_w$ and $T_a$ denote the waiting time and the time slot during which the update comes.
	\end{IEEEproof}
\begin{remark}
	\label{remark_comp_l1l2}
It is difficult to tell from (\ref{eq_MeanPAoI_l1}) and (\ref{eq_MeanPAoI_l2}) which load model provides a lower mean PAoI since the PAoI in (\ref{eq_MeanPAoI_l2}) starts at a lower value but has one more waiting time term. However, we can expect that Load Model 2 performs better since the value of $X_n$ is at most $N_d$ which equals to the minimum value of PAoI in (\ref{eq_MeanPAoI_l2}), and this performance gap (gap in mean PAoI) depends on the arrival rate and increases with the increasing of $N_d\times T$.
\end{remark}

\subsubsection{Uncorrelated Devices}
Compared to the correlated devices, the mean PAoI of uncorrelated devices also includes one more successful transmission, as shown in the following theorem.
\begin{theorem}[Mean PAoI under Load Model 2 and Uncorrelated Devices]  In the case of uncorrelated devices, the mean PAoI under the Load Model 2 is given by
	\begin{align}
		\bar{\Delta}_{l_2} &= \mathbb{E}_{R_u}[\bar{F}_{\Delta_{o}^{'}}(n\times T)+T_{n}\times T] + X_n,
	\end{align}
	in which $T_n$ denotes the mean success transmission time,
	\begin{align}
		T_{n} &= \sum_{t_n = 0} (t_n N_d+1)P_s(R_u)(1-P_s(R_u))^{t_n}.
	\end{align}
\end{theorem}

Now we have obtained all the required equations and in the next section we validate all these results with simulations.

\section{Numerical Results}
\label{sect_numerical}
In this section, we validate the analytical results with simulations, evaluate the impact of number of devices, UAV altitudes, and environment parameters on the mean PAoI, and compare the two load models. Unless stated otherwise, we use the simulation parameters as listed herein Table \ref{par_val}.

\begin{table}\caption{Table of Parameters}\label{par_val}
	\vspace{-4mm}
	\centering
	\begin{center}
		\resizebox{1\columnwidth}{!}{
			\renewcommand{\arraystretch}{1}
			\begin{tabular}{ {c} | {c} | {c}  }
				\hline
				\hline
				\textbf{Parameter} & \textbf{Symbol} & \textbf{Simulation Value}  \\ \hline
				Density of IoT clusters & $\lambda_{ u}$ & $1$ km$^{-2}$ \\\hline
				MCP disk radius & $r_c$ & 120 m \\\hline
				UAV altitude & $h$ & 100 m \\\hline
				Environment parameters (highrise urban area) \cite{al2014optimal} & $ (a,b)$ & $(27,0.08)$ \\\hline				
				Environment parameters (dense urban  area) \cite{al2014optimal} & $ (a,b)$ & $(12,0.11)$ \\\hline
				Environment parameters (suburban urban  area) \cite{al2014optimal} & $ (a,b)$ & $(9.6,0.16)$ \\\hline
				Environment parameters (urban area) \cite{al2014optimal} & $ (a,b)$ & $(4.88,0.43)$ \\\hline
				Power control parameters & $\rho_{\{l,n\}}$ & 0.001 W, 0.001 W\\\hline
				Maximum transmission power &  $p_{u}$ &  0.1 W\\\hline
				Noise power & $\sigma^2 $ & $10^{-9}$ W\\\hline
				SINR threshold & $\theta$ & $0$ dB \\\hline
				N/LoS path-loss exponent & $\alpha_{ n},\alpha_{ l}$ & $4,2.1$ \\\hline
				N/LoS fading gain & $m_{ n},m_{ l}$ & $1,3$ \\\hline
				N/LoS additional loss& $\eta_{ n},\eta_{ l}$ & $-20,0$ dB 
				\\\hline\hline
		\end{tabular}}
	\end{center}
	\vspace{-4mm}
\end{table}

In the simulations of the considered system setup, we first generate one realization of PPP to model the locations of IoT cluster centers and the typical IoT cluster is centered at the origin.  For the typical IoT cluster,  we generate the locations of $N_d $ IoT devices which are randomly located within the cluster following a uniform distribution. For the remaining IoT clusters, we generate one IoT device for each cluster to model the interferer. For the typical IoT cluster, we first compute the success probability for each device and then compute the mean PAoI under the two load models. Besides, the unit of mean PAoI in the following figures is $T$.

\begin{figure}
	\centering
	\subfigure[]{\includegraphics[width = 0.8\columnwidth]{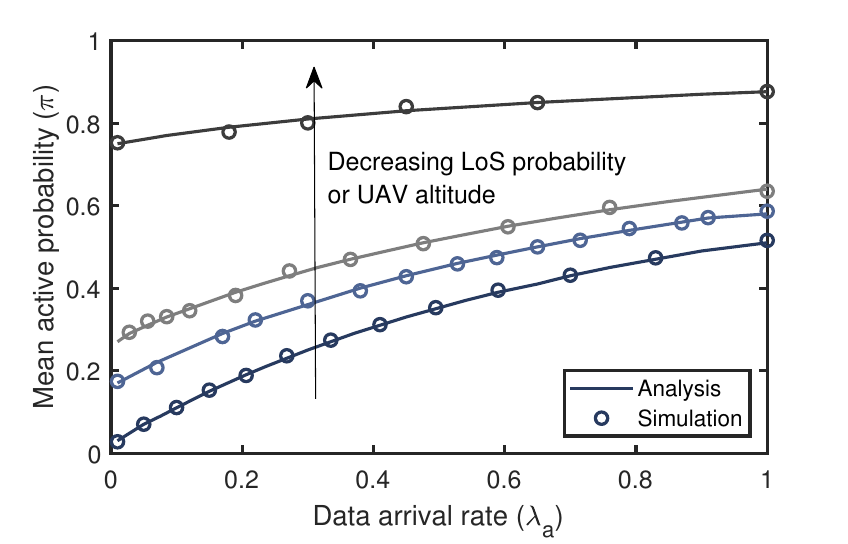}}
	\subfigure[]{\includegraphics[width = 0.8\columnwidth]{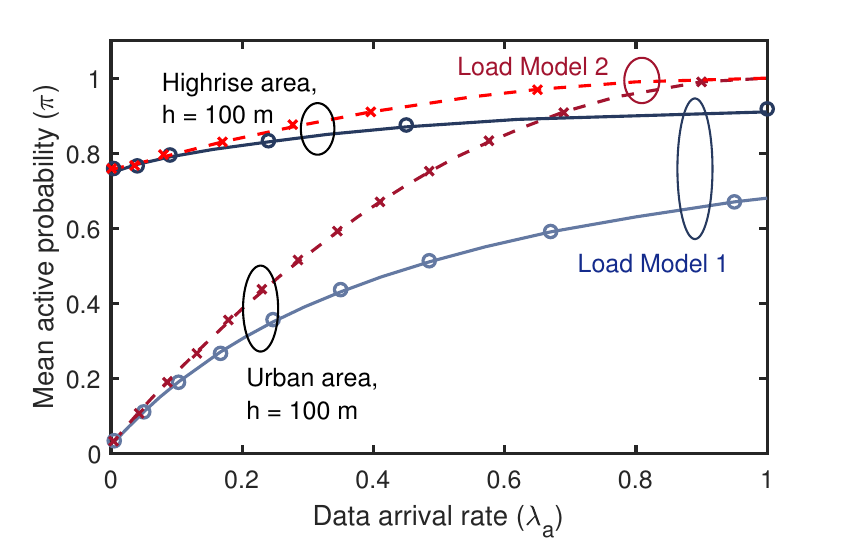}}	
	\caption{Simulation and analytical results of computing the mean activity probability of IoT devices in the case of \textbf{(a)} $N_d = 1\times T$, these four curves, from the bottom to the top, are plotted under (1) suburban urban  area and $h = 100$ m, (2) urban area and $h = 100$ m, (3) urban area and $h = 80$ m, and highrise urban  area and $h = 100$ m; \textbf{(b)} Load Model 1 and Load Model 2, and $N_d = 2\times T$.}
	\label{fig_active}
\end{figure}
We first show the accuracy of computing the mean activity probability,  as shown in Fig. \ref{fig_active} (a), in which four curves, from the bottom to the top, are plotted under (1) suburban area and $h = 100$ m, (2) urban area and $h = 100$ m, (3) urban area and $h = 80$ m, and highrise urban  area and $h = 100$ m. The simulation results show a good matching with the analytical results, which is obtained by iterating as mentioned in Theorem \ref{theorem_activeprob}. We observe that activity probability increases with the increase of information arrival rate (due to more data generated) and increases with the decrease of LoS probability (due to longer successful transmission time). We then compare the activity probability under Load Model 1 and Load Model 2, as shown in Fig. \ref{fig_active} (b), in which we plot two environments (1) highrise urban  area and $h = 100$ m, and (2) urban area and $h = 100$ m.
Interestingly, for Load Model 1, as the arrival rate increases $0$ to $1$, the activity probability has a smaller range and it cannot reach $1$. This is because of the state transfer diagram shown in Fig. \ref{fig_ill_stead}: it takes at least one time slot for devices to generate the data, hence, the activity probability approaches $1$ with the decrease of success probability or the increase of transmission time (increases of $N_d$). For Load Model 2, it has a higher interference than Load Model 1 and the devices activity probability is able to reach $1$. This is because the UAV uses the data generation time to serve other devices.

\begin{figure}
	\centering
	\subfigure[]{\includegraphics[width = 0.8\columnwidth]{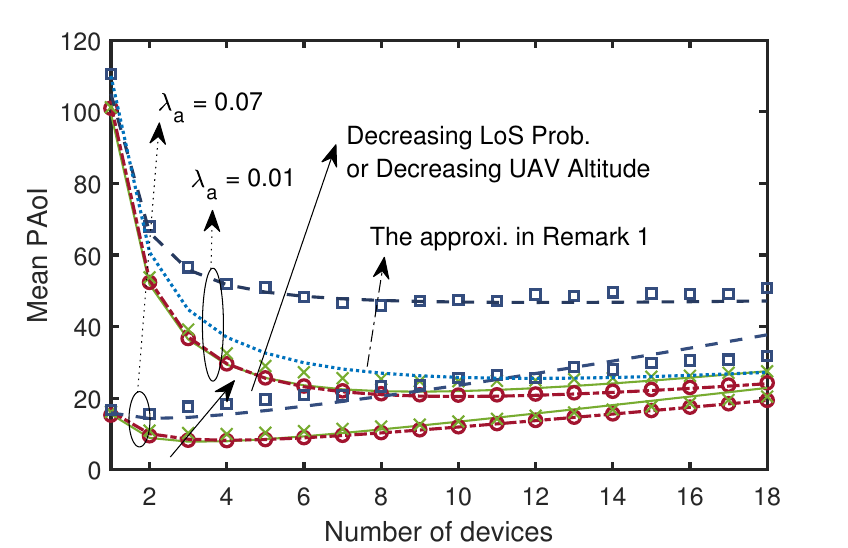}}
	\subfigure[]{\includegraphics[width = 0.8\columnwidth]{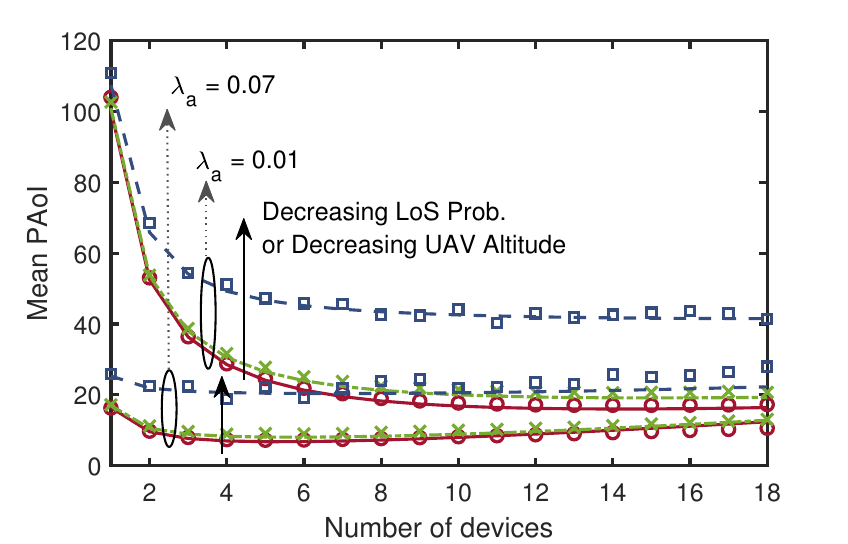}}	
	\caption{Simulation and analysis results of the mean PAoI of correlated devices: \textbf{(a)} Load Model 1, split the bandwidth, \textbf{(b)} Load Model 2, split the time.}
	\label{fig_dif_par_com}
\end{figure}
In Fig. \ref{fig_dif_par_com}, we plot the mean PAoI under two load models: Fig. \ref{fig_dif_par_com} (a) bandwidth splitting and  Fig. \ref{fig_dif_par_com} (b) time splitting, while the solid lines with crossing markers denote the mean PAoI under suburban areas and the dash lines with circle markers denotes the mean PAoI under dense areas and UAV altitudes $h = 80$ m (we omit the curve under dense area and $h = 100$ since these three curves are very close to each other). Moreover, the approximation mentioned in Remark \ref{remark_app} shows good matching in these scenarios, thus, is omitted in the figure. The mean PAoI only slightly increases when the environment changes from suburban to dense areas and UAV altitude decreases. This is because, in the uplink transmission of UAV-assisted networks, the success probability decreases slowly when the transmission distance increases, and since we consider that devices are spatially clustered the transmission distance is limited by the cluster radii. However, the mean PAoI increases dramatically under highrise urban  areas (dash lines with square markers), and the approximation mentioned in Remark \ref{remark_app} shows gaps (Fig. \ref{fig_dif_par_com} (a) dotted line). This is because, in highrise urban  areas, the LoS probability and the success probability decrease dramatically when the transmission distance increases. Thus, the devices near the cluster center have a much higher success probability compared to the devices at the edge, and considering all the devices have the same success probability is not accurate: while the devices near the cluster center update the information periodically, the devices at the edge rarely transmit useful information.

\begin{figure}
	\centering
	\subfigure[]{\includegraphics[width = 0.8\columnwidth]{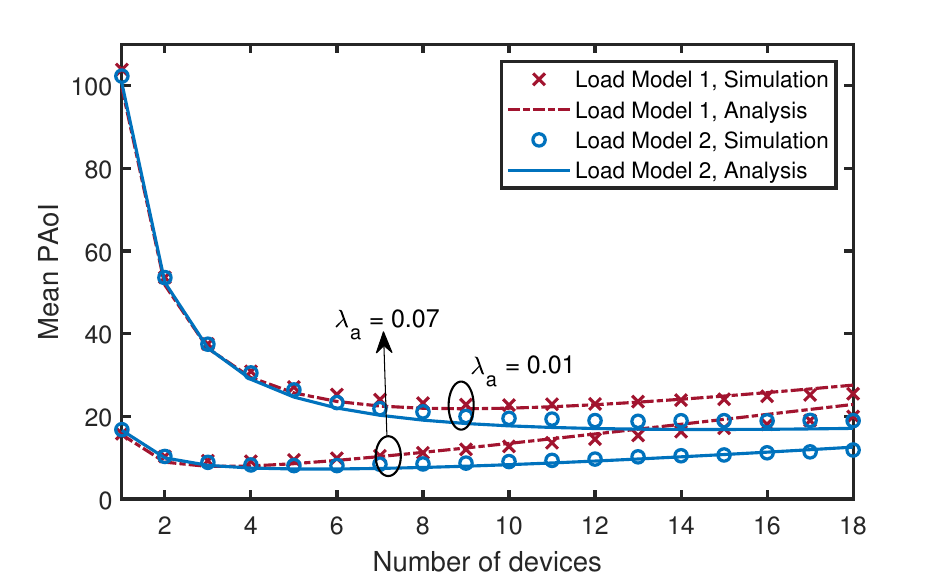}}
	\subfigure[]{\includegraphics[width = 0.8\columnwidth]{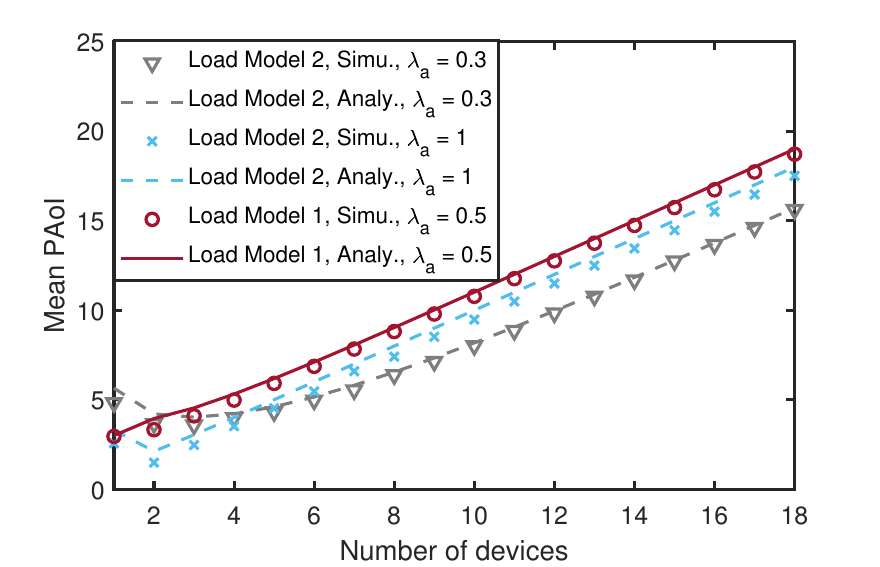}}	
	\caption{Simulation and analysis results of the mean PAoI of correlated devices: \textbf{(a)} mean PAoI comparison between Load Model 1 and Load Model 2, \textbf{(b)} at different arrival rates of Load Model 2.}
	\label{fig_splitBT_comp}
\end{figure}
In Fig. \ref{fig_splitBT_comp} (a), we compare mean PAoI under the proposed two load models in the case of correlated devices (dense areas with altitudes $h = 100$ m). We observe that an optimal number of devices exists for a fixed data arrival rate and Load Model 2 performs slightly better than Load Model 1 at a large number of devices, as expected in Remark \ref{remark_comp_l1l2}. Mean PAoI decreases first due to a higher data arrival rate at the UAV side, and increases due to a longer transmission time in Load Model 1 and a longer waiting time in Load Model 2. In addition, Load Model 2 schedules the transmission of updates since we consider UAVs serving the devices periodically. The update information comes when the previous update is successfully transmitted. Hence, the updates are scheduled since they also come periodically. However, we would like to indicate that these two load models are fairly compared since all the devices are fairly served. Interestingly, we notice that while the optimal value of the number of devices only exists at a low value of arrival rate in Load Model 1 ($\lambda_a \leq 0.5$, as shown in Fig. \ref{fig_splitBT_comp} (b)) under dense area and $h = 100$ m, optimal values exist (number of devices equals to $2$) for all the arrival rates in Load Model 2, see Fig. \ref{fig_splitBT_comp} (b). This is because we consider that devices cannot have any updates during the transmission period. For instance, when the $\lambda_a =1$ and $P_s = 1$, the mean PAoI is $3\times T$ which contains two successful transmissions and one data update time slot in the case of a single device, and the mean PAoI is $2$ which contains one successful transmission and one data update time slots in the case of two devices. Hence, serving two correlated devices is more efficient than serving one device, due to the efficient use of the data generation time slot to collect data from another device.

\begin{figure}
	\centering
	\subfigure[]{\includegraphics[width = 0.8\columnwidth]{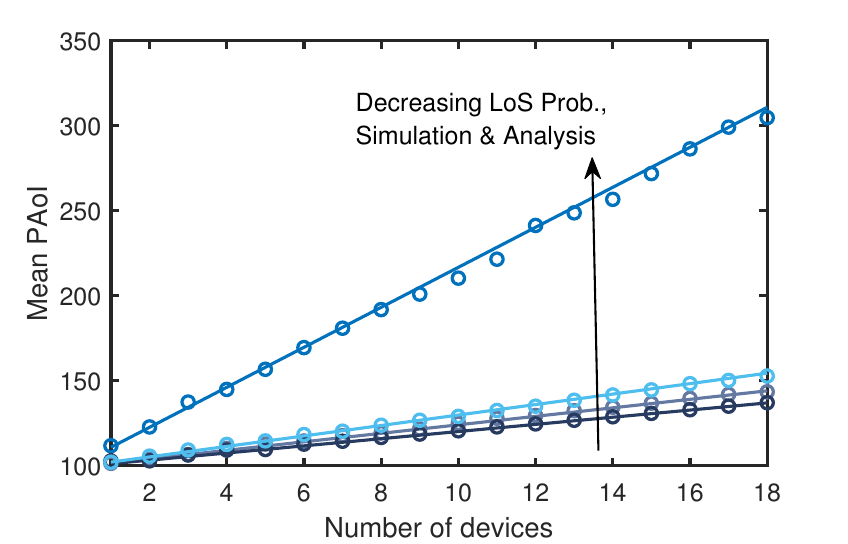}}
	\subfigure[]{\includegraphics[width = 0.8\columnwidth]{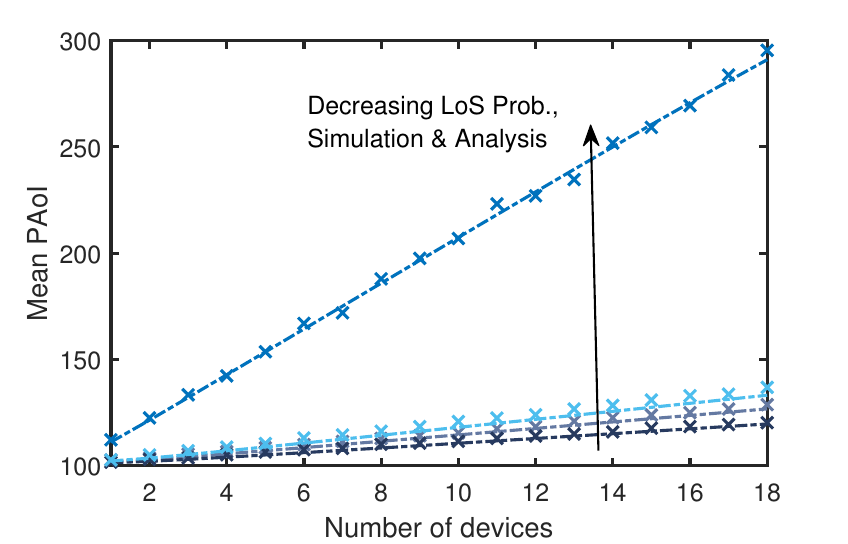}}	
	\caption{Simulation and analysis results of the mean PAoI of uncorrelated devices: \textbf{(a)} mean PAoI under Load Model 1, \textbf{(b)} mean PAoI under Load Model 2. In both \textbf{(a)} and \textbf{(b)}, the curves from the top to the bottom are plotted under: highrise urban  areas $a = 27,b = 0.08$ and $h = 100$, dense area $a = 12,b = 0.16$ and $h = 80$, dense area $a = 12,b = 0.16$ and $h = 100$, and suburban area $a = 4.88,b = 0.43$ and $h = 100$.}
	\label{fig_unco_dif_par}
\end{figure}

In Fig. \ref{fig_unco_dif_par}, we plot the mean PAoI under the two load models in the case of uncorrelated devices. For uncorrelated devices, the mean PAoI increases with the increase in the number of devices. The mean PAoI keeps almost the same when the UAV altitudes change from $h = 100$ to $h = 80$ and the environment changes from suburban to dense areas, while the mean PAoI increases dramatically at highrise urban  areas. This is because the coverage probability (the expectation of the success probability over the transmission distance) under the highrise urban  area is about $0.24$ while it is about $0.6$ to $0.9$ in the three scenarios. Thus, it takes a much longer time to successfully transmit the update in highrise urban  areas.

\begin{figure}
	\centering
	\subfigure[]{\includegraphics[width = 0.8\columnwidth]{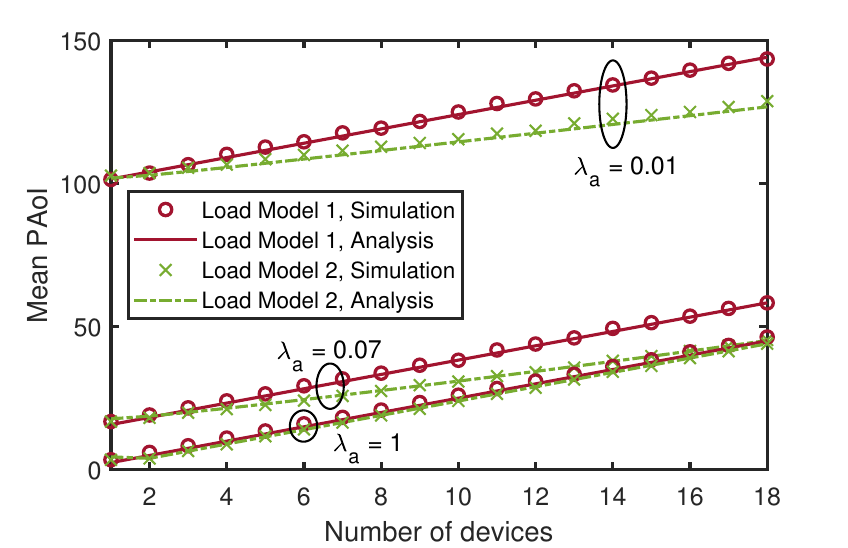}}
	\subfigure[]{\includegraphics[width = 0.8\columnwidth]{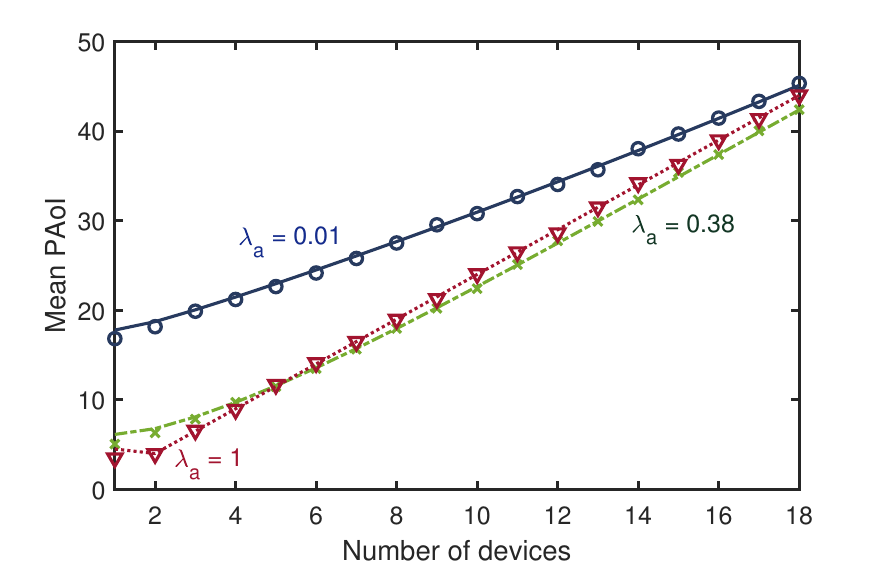}}	
	\caption{Simulation and analysis results of the mean PAoI of uncorrelated devices: \textbf{(a)} mean PAoI comparison between Load Model 1 and Load Model 2, \textbf{(b)} at different arrival rates of Load Model 2.}
	\label{fig_unco_comp_splitBT}
\end{figure}
In Fig. \ref{fig_unco_comp_splitBT} (a), we plot the mean PAoI of the proposed two load models in the case of uncorrelated devices. Similar to the trend observed in Fig. \ref{fig_splitBT_comp} (a), Load Model 2 performs slightly better than Load Model 1, as expected in Remark \ref{remark_comp_l1l2}.  Additionally, we notice that the mean PAoI gap decreases with the increasing of arrival rate, e.g., at a low arrival rate the gap is about $2(N_d-2-X_n)$ and at a high arrival rate ($\lambda_{a} = 1$) the gap is about $1$.

 To explain this, we consider the scenario that $N_d = 18$, $\lambda_{a,1} = 1$ and $\lambda_{a,2} = 0.2$. When  $\lambda_{a,1} = 1$, the data updates at the first time slot in both load models, and the device starts to transmit the update in the following $T_{m1} = 18$ time slots in Load Model 1 while the device waits for $T_{m2}-2 = 16$ time slots and transmits in the $18$-th time slot in Load Model 2. If both transmissions succeed, the Load Model 2 requires $1$ less time slot. If the transmissions fail, both devices spend another $N_d$ to re-transmit. Hence, the Load Model 2 still requires $1$ less time slot. In the case of $\lambda_{a,2} = 0.2$, the Load Model 1 device transmits the update as soon as it comes, and intuitively, it should perform better. However, the minimum value of AoI of Load Model 2 is lower when it arrives at the serving UAV since the update may come in the middle of a period, e.g., coming $1$ time slot before the transmission, hence, the AoI is $2\times T$ when it arrives at the UAV. Besides, the inter-arrival time is also longer in Load Model 1: new updates come when the previous updates are successfully transmitted. Therefore, the inter-arrival time in Load Model 1 is $T_{m1}+\frac{1}{\lambda_a}$ while in load model 2 is $T+\frac{1}{\lambda_a}$, in which $T_{m1} = N_d T$.

Besides, we notice in Fig. \ref{fig_unco_comp_splitBT} (b) that in Load Model 2, a higher arrival rate does not always result in a lower mean PAoI: for instance, we plot $\lambda_{a,1} = 1$ and $\lambda_{a,2} = 0.38$ in Fig. \ref{fig_unco_comp_splitBT} (b). When the number of devices is smaller than $N_d \leq 5$, mean PAoI under $\lambda_{a,1}$ is smaller while $N_d >5$, $\lambda_{a,2}$ has a smaller mean PAoI. This is because of the waiting time $X_n$. At high arrival rates, updates have a higher probability of arriving at the beginning of the period and waiting for a longer time.

\section{Conclusion}

This work presented a stochastic geometry-based analysis of PAoI performance metric in a UAV-assisted IoT network in which we considered correlated IoT devices and uncorrelated devices and two load models, assuming a non-preemptive queuing discipline. We first showed the impact of UAV altitudes and environment parameters on the mean PAoI. The mean PAoI increases dramatically in high dense areas, while keeps almost the same in suburban to dense areas and slight changes in UAV altitudes. We then compared the mean PAoI under two load models. For the correlated devices, we showed that an optimal value exists to minimize the mean PAoI when the arrival rate is low in Load Model 1, and two devices always have a lower PAoI in Load Model 2 for all arrival rates. In particular, we showed that even though Load Model 2, time splitting, causes a higher interference, it provides a lower mean PAoI than Load Model 1, bandwidth splitting, for both scenarios for IoT devices. This is because of the spatially clustered devices and the  higher probability of establishing LoS links. Additionally, it is more time efficient for UAVs to split the time to serve multiple devices as it can use data generation time to serve other devices.

 In future research, an interesting avenue to explore would be the analysis of AoI in mobility-aware IoT networks. For instance, studying scenarios where UAVs dynamically move around to collect updates from IoT devices belonging to different IoT clusters or where IoT devices switch their connections between different UAVs as they move.
Additionally, investigating the impact of limited resources on UAVs' AoI is another relevant direction for future work. UAVs possess finite resources, including battery power, processing capability, and communication bandwidth. Optimizing resource allocation to minimize PAoI, while considering these limitations poses a challenging task that could be tackled in further research. 

\appendix

\subsection{The Laplace Transform in Lemma \ref{lemma_Mb}}\label{app_laplace}
The expression of the Laplace transform in Lemma \ref{lemma_Mb} is provided in \cite[Lemma 2]{Upref} and given by
\begin{strip}
	\begin{align*}
	&\mathcal{L}(g_{1}(r),\cdots,g_{b}(r)) =  \exp(-(g_{1}(R_u)+\cdots+g_{b}(R_u))\sigma^2)\nonumber\\
	& \exp\bigg(-2\pi\bar{\pi}\lambda_u\int_{h}^{\infty}\int_{h}^{\sqrt{h^2+r_c^2}}\bigg[1-\bigg(\frac{m_l}{m_l+g_{1}(R_u)\rho_l\eta_{ l}  r^{\alpha_{l}\epsilon_l}z^{-\alpha_{l}}}\cdots\frac{m_l}{m_l+g_{b}(R_u)\rho_l \eta_{ l} r^{\alpha_{l}\epsilon_l}z^{-\alpha_{l}}}\bigg)^{m_l}\bigg]\nonumber\\
	& \times z P_l(\sqrt{r^2-h^2}) P_l(\sqrt{z^2-h^2})f_{\rm R_u}(r){\rm d}r{\rm d}z\bigg)\nonumber\\
	& \exp\bigg(-2\pi\bar{\pi}\lambda_u\int_{h}^{\infty}\int_{h}^{\sqrt{h^2+r_c^2}}\bigg[1-\bigg(\frac{m_l}{m_l+g_{1}(R_u)\rho_n\eta_{ l} r^{\alpha_{n}\epsilon_n}z^{-\alpha_{l}}}\cdots\frac{m_l}{m_l+g_{b}(R_u)\rho_n\eta_{ l} r^{\alpha_{n}\epsilon_n}z^{-\alpha_{l}}}\bigg)^{m_l}\bigg]\nonumber\\
	& \times z P_n(\sqrt{r^2-h^2}) P_l(\sqrt{z^2-h^2})f_{\rm R_u}(r){\rm d}r{\rm d}z\bigg)\nonumber\\
	&\exp\bigg(-2\pi\bar{\pi}\lambda_u\int_{h}^{\infty}\int_{h}^{\sqrt{h^2+r_c^2}}\bigg[1-\bigg(\frac{m_n}{m_n+g_{1}(R_u)\rho_l \eta_{ n} r^{\alpha_{l}\epsilon_l}z^{-\alpha_{n}}}\cdots\frac{m_n}{m_n+g_{b}(R_u)\rho_l \eta_{ n} r^{\alpha_{l}\epsilon_l}z^{-\alpha_{n}}}\bigg)^{m_n}\bigg]\nonumber\\
	&\times z P_l(\sqrt{r^2-h^2}) P_n(\sqrt{z^2-h^2})f_{\rm R_u}(r){\rm d}r{\rm d}z\bigg)\nonumber\\
	&\exp\bigg(-2\pi\bar{\pi}\lambda_u\int_{h}^{\infty}\int_{h}^{\sqrt{h^2+r_c^2}}\bigg[1-\bigg(\frac{m_n}{m_n+g_{1}(R_u)\rho_n\eta_{ n} r^{\alpha_{n}\epsilon_n}z^{-\alpha_{n}}}\cdots\frac{m_n}{m_n+g_{b}(R_u)\rho_n\eta_{n} r^{\alpha_{n}\epsilon_n}z^{-\alpha_{n}}}\bigg)^{m_n}\bigg]\nonumber\\
	&\times z P_n(\sqrt{r^2-h^2}) P_n(\sqrt{z^2-h^2})f_{\rm R_u}(r){\rm d}r{\rm d}z\bigg),\label{app_laplacetransform_eq}
\end{align*}
\end{strip}
	 in which $g_{l,i}(r) = k_i\beta_2(m_l)m_l\theta r^{(1-\epsilon_{l})\alpha_{ l}}(\rho_l\eta_{ l})^{-1}$,  $g_{n,i}(r) = k_i\beta_2(m_n)m_n\theta r^{(1-\epsilon_{n})\alpha_{ n}}(\rho_n\eta_{ n})^{-1}$, and $\mathcal{L}(g_{1}(r),\cdots,g_{b}(r))$, $\beta_2(m) = (m!)^{-1/m}$ and  $\bar{\pi}$ is the mean activity probability.

\bibliographystyle{IEEEtran}
\bibliography{Ref14}

\end{document}